\documentclass[fleqn,twoside,twocolumn,nofootinbib]{revtex4} % Specifies the document class %,unsortedaddress
\usepackage{ujp} % \usepackage[cyr]{ujp} for cyrillic, \usepackage[web]{ujp} for web
\begin{document}
\title[INFLUENCE OF $\gamma$-IRRADIATION]%колонтитул
{INFLUENCE OF \boldmath$\gamma$-IRRADIATION ON INITIAL\\ MAGNETIC
PERMEABILITY OF
AMORPHOUS\\ AND NANOCRYSTALLINE Fe--Si--B-BASED ALLOYS}%
\author{V.Yu.~Povarchuk}%1 автор
\affiliation{Institute of Physics, Nat. Acad. of Sci. of
Ukraine}%институт
\address{46, Prosp. Nauky, Kyiv 03028, Ukrain}%адрес
\email{pvyu@i.ua}%e-mail
\author{V.K.~Nosenko}%
\affiliation{G.V.~Kurdyumov Institute for Metal Physics, Nat. Acad.
of Sci.
of Ukraine }%
\address{36, Academician Vernadskyi Ave., Kyiv 03142, Ukraine}%
\author{A.M.~Kraitchinskii}
\affiliation{Institute of Physics, Nat. Acad. of Sci. of
Ukraine}%институт
\address{46, Prosp. Nauky, Kyiv 03028, Ukrain}%адрес
\email{pvyu@i.ua}%e-mail
\author{V.B.~Neimash}
\affiliation{Institute of Physics, Nat. Acad. of Sci. of
Ukraine}%институт
\address{46, Prosp. Nauky, Kyiv 03028, Ukrain}%адрес
\email{pvyu@i.ua}%e-mail
\author{M.M.~Kras'ko}
\affiliation{Institute of Physics, Nat. Acad. of Sci. of
Ukraine}%институт
\address{46, Prosp. Nauky, Kyiv 03028, Ukrain}%адрес
\email{pvyu@i.ua}%e-mail
\author{\fbox{V.V.~Maslov}}
\affiliation{G.V.~Kurdyumov Institute for Metal Physics, Nat. Acad.
of Sci. of Ukraine }%
\address{36, Academician Vernadskyi Ave., Kyiv 03142, Ukraine}%

 \udk{???} 
\razd{\secix}
\setcounter{page}{345}%
\maketitle

%\makeatletter
%\renewcommand{\thesection}{\arabic{section}}
%\renewcommand{\p@subsection}{}
%\renewcommand{\thesubsection}{\arabic{section}.\arabic{subsection}}
%\renewcommand{\p@subsubsection}{}
%\renewcommand{\thesubsubsection}
%{\arabic{section}.\arabic{subsection}.\arabic{subsubsection}}
%\makeatother

%\input{tcilatex}

\begin{abstract}
By determining the inductance factor, the dependence of the initial
magnetic permeability $\mu _{i}$ of amorphous and nanocrystalline
Fe--Si--B-based alloys on the $\gamma $-irradiation dose has been
studied. The doping of amorphous Fe--Si--B alloys with nickel and
molybdenum is found to enhance the radiation sensitivity of $\mu
_{i}$. The initial magnetic permeability of nanocrystalline magnetic
alloys is determined to be less sensitive to the action of $\gamma
$-radiation than that of doped amorphous ones. A hypothesis is
put forward that the influence of radiation on the initial magnetic
permeability is associated with the creation of non-magnetic
inclusions in the structure of amorphous alloys and in the amorphous
matrix of nanocrystalline alloys.
\end{abstract}

\section{Introduction}

Amorphous and nanocrystalline alloys on the basis of the Fe--Si--B
system are characterized by high values of saturation induction and
magnetic permeability and by low magnetic reversal losses.
Therefore, they are widely applied in the manufacture of inductive
elements for the electrotechnical equipment: power transformers,
small-sized transformers operating at high frequencies, throttles, {\it etc}.
From the scientific viewpoint, the researches of the radiation influence
on structurally sensitive characteristics of amorphous and
nanocrystalline alloys can be useful for establishing the mechanisms
of radiation-induced structural changes in such systems. An
application purpose of our research was to study the expediency of
using amorphous and nanocrystalline ferromagnets on the basis of
the Fe--Si--B system under ionizing radiation conditions, as well as a
possibility to control the properties of those substances taking
advantage of their radiation treatment.\looseness=1

Available literary data did not enable us to estimate unambiguously
the influence of $\gamma $-radiation on magnetic properties of
amorphous soft magnetic alloys. The authors of work \cite{1}
reported that the irradiation to the dose $\Phi
=10^{9}~\mathrm{rad}$ stimulates a growth of the initial magnetic
permeability $\mu _{i}$ and the residual induction $B_{r}$, as well
as a reduction of $H_{c}$ for magnetic circuits fabricated from
amorphous soft magnetic alloys Fe$_{85-x}$Co$_{x}$B$_{15}$
($x=12\div 25{\%}$). On the contrary, the results of work \cite{2}
testified to a deterioration of magnetic properties of amorphous
FeNiMoSiB alloys subjected to this kind of radiation treatment. The
irradiation of initial (nonannealed) amorphous alloy
Co$_{83.5}$Fe$_{5.5} $Si$_{8.5}$B$_{2.5}$ to the dose $\Phi
=10^{7}~\mathrm{rad}$ stimulated the growth of both $B_{r}/B_{s}$
and $H_{c}$ \cite{3}. For today, there are no data concerning the
influence of $\gamma $-radiation on magnetic properties of
nanocrystalline alloys on the basis of \mbox{Fe--Si--B}.\looseness=1

Ambiguous data on the influence of $\gamma $-radiation on magnetic
characteristics of amorphous alloys made it expedient to study their dose
dependences in the range as wide as possible. Evidences in favor of this
purpose are also given by a nonmonotonous character of the dose dependence
revealed by the first maximum height in the structural factor \cite{4}.

\section{Experimental Materials, Methods of Their Fabrication, and Research
Techniques}

In this work, we studied the influence of $\gamma $-radiation
($^{60}$Co was a radiation source, $E_{\gamma }=1.17$ and
1.33~$\mathrm{~MeV}$, the flow density $\phi _{\gamma }\approx
10^{11}\gamma
_{\mathrm{quant}}/(\mathrm{cm}^{\mathrm{2}}\mathrm{s})$) applied in
doses up to $5.7\times 10^{18}~\mathrm{cm}^{-2}=3.4\times
10^{9}~\mathrm{rad}$ on the initial magnetic permeability of
amorphous alloys of the Metglas type (MG-alloys) and nanocrystalline
alloys of the Finemet type (FM-alloys). The chemical compositions of
MG-alloys are given in Table~1 and those of FM-alloys in Table~2. From
Table~1, one can see that the MG-alloys can be divided into two groups
according to the concentration of nonmetallic components in them:
Fe-Si$_{6}$B$_{14}$ (alloys MG{\-}1 to MG-3) and
Fe--Si$_{2}$B$_{16}$ (alloys MG-5 to MG-8). All alloys, but MG-1 and
MG-5, were doped with Ni and Mo atoms. Doping MG-alloys with those
atoms enhances their thermal stability and improves their magnetic
properties~\cite{5}.\looseness=1

%Таблиця. 1.
\begin{table}[b] \noindent\caption{Chemical composition (at.\%)
 of amorphous alloys of the Metglas type }
\vskip3mm\tabcolsep10.6pt

\noindent{\footnotesize\begin{tabular}{c c c c c c c}
 \hline \multicolumn{1}{c}
{\rule{0pt}{9pt}No} & \multicolumn{1}{|c}{Alloy}&
\multicolumn{1}{|c}{Fe}& \multicolumn{1}{|c}{Si}&
\multicolumn{1}{|c}{B}& \multicolumn{1}{|c}{Ni}&
\multicolumn{1}{|c}{Mo}\\%
\hline%
1&{MG-1}& 80& 6& 14& &\\
2&{MG-2}& 76.2& 6& 14& 3.8&\\
3&{MG-3T}& 78.5& 6& 14& 1&0.5 \\
4&{MG-5}& 82& 2& 16& &\\
5&{MG-6}& 78& 2& 16& 1&3 \\
6&{MG-7}& 77.5& 2& 16& 3.5&1 \\
7&{MG-8}& 75.5& 2& 16& 3.5&3 \\
\hline
\end{tabular}}
\end{table}

%Таблиця 2.

\begin{table}[b]
\noindent\caption{Chemical composition (at.\%) of amorphous alloys
of the Finemet type}\vskip3mm\tabcolsep7.2pt

\noindent{\footnotesize\begin{tabular}{c c c c c c c c}
 \hline \multicolumn{1}{c}
{\rule{0pt}{9pt}No} & \multicolumn{1}{|c}{Alloy}&
\multicolumn{1}{|c}{Fe}& \multicolumn{1}{|c}{Si}&
\multicolumn{1}{|c}{B}& \multicolumn{1}{|c}{Cu}&
\multicolumn{1}{|c}{Nb}&
\multicolumn{1}{|c}{Co}\\%
\hline%
1&FM-2T&73&15.8&7.2&1&3\\
2&FM-6&73.6&15.8&7.2&1&2.4\\
3&FM-10&71.25&16.4&7.7&1&2.1&1.55\\
4&FM-11&70.05&16.4&9&1&2&1.55\\
\hline
\end{tabular}}
\end{table}

For the preparation of initial alloys, components of high purity
were used: Fe -- 99.96~wt.\%, Si -- 99.999~wt.\%, B -- 99.9~wt.\%,
Ni $\geq $ 99.9~wt.\%, Mo $\geq $ 99.8~wt.\%, Cu $\geq $ 99.9~wt.\%,
Nb $\geq $ 99.9~wt.\%, and Co $\geq $ 99.9~wt.\%. Alloy MG-3T was
produced from components of technical purity: Fe $\geq $
98.6~wt.\%, Mo~$\geq $~99~wt.\%, ferroboron FB-17 (B $\approx $
20.5~wt.\%, Al -- 1.2{\%}, Si -- 0.4{\%}), and ferronickel FN-5 (Ni
-- 7.8~wt.\%, Co -- 0.36{\%}, Cu -- 0.11{\%}, Al -- 0.1{\%}, Si --
0.03{\%}).

Alloys were produced in an induction furnace, in an inert Ar atmosphere. The
chemical composition of alloys was determined using the method of X-ray
fluorescence analysis. Amorphous ribbons 10~mm in width and 24--26~$\mu
\mathrm{m}$ in thickness were fabricated using the method of melt spinning
in air on an open-type installation.

To study the magnetic properties of amorphous and nanocrystalline
alloys, the initial (nonannealed) amorphous ribbon was used for the
fabrication of magnetic circuits (circular cores) with the help of a
special equipment for winding with the lowest tension. The
geometrical parameters of twisted circular cores in the specimens
were as follows: an internal diameter of 12~mm, an external diameter
of 20~mm, and a height of 10~mm. Magnetic circuits made of MG-alloys
were annealed at a temperature of 420$~^{\circ }\mathrm{C}$ for
15~min in the carbon dioxide atmosphere. Cores fabricated from FM-6,
FM-10, and FM-11 alloys were annealed in the carbon dioxide
atmosphere at the temperature $T=520~^{\circ }\mathrm{C}$ for
30~min, and those from alloy FM-2T at $T=535~^{\circ }\mathrm{C}$
for 60~min. To measure the initial magnetic permeability
($H_{m}=0.2$~A/m, $f=10$ and 100~kHz), we applied the method of
inductance factor determination \cite{6}. The determination accuracy
for $\mu _{i}$ was~5\%.\looseness=1

\section{Experimental Results and Their Discussion}

The thermal treatment of MG-alloys gave rise to an increase of their initial
magnetic permeabilities $\mu _{i}$ by several times and a reduction of the
corresponding coercive forces $H_{c}$; the hysteresis loop became narrower,
and the number of Barkhausen jumps diminished. Those phenomena are connected
with a reduction of the magnetic anisotropy owing to the elimination of internal
stresses \cite{7}. These alloys find a wide application in the industry. The
magnetic characteristic, which is the most sensitive to their structure, is
the initial magnetic permeability $\mu _{i}$. In this work, we report
results of our researches concerning the influence of radiation on this
parameter obtained for magnetic circuits annealed in the optimum regime.

%Fig. 1
\begin{figure*}% figure* for wide figure, [h] [!] to change the placement
\includegraphics[width=8.5cm]{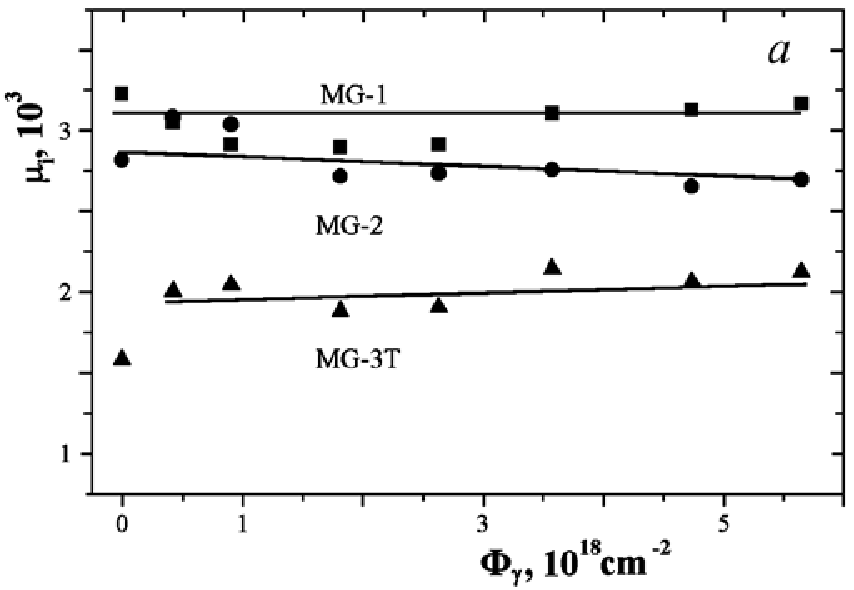}\hspace{0.5cm}\includegraphics[width=8.5cm]{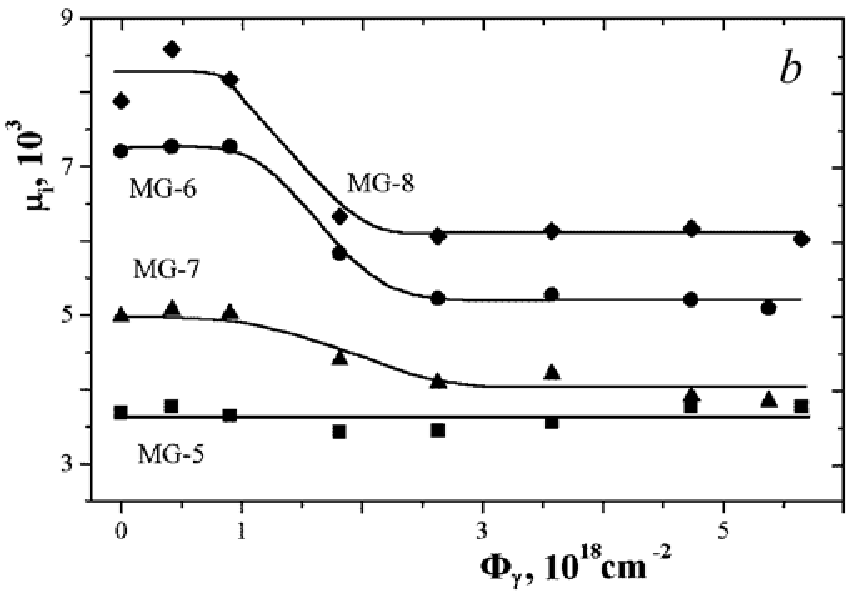}
%\vskip-3mm
\caption{Dose dependences of the initial magnetic permeability
($f=10~\mathrm{kHz} $) of magnetic circuits fabricated from
MG-alloys  }
\end{figure*}

From Fig.~1, one can see that radiation did not affect the initial magnetic
permeability of undoped alloys MG-1 and MG-5. In MG-alloys doped with nickel
and molybdenum, a reduction of $\mu_{i}$ was observed, as the exposure dose
of $\gamma$-radiation grew. The largest relative variations of $\mu_{i}$
were observed for alloys MG-6 and MG-8 (by 28 and 23\%, respectively), in
which the Ni concentration amounted to 1 and 3.5\%, respectively, and the Mo
concentration was 3\%. For alloy MG-7, in which the Ni concentration was
3.5\% and the Mo concentration was 1\%, the relative variation of $\mu_{i}$
was about 20\%. An insignificant reduction of the initial magnetic permeability
was observed for alloy MG-2, which contained 3.8\% Ni and did not contain
Mo. Hence, the reduction of the initial magnetic permeability in the examined
MG-alloys under the action of $\gamma$-radiation can be associated, first of
all, with the presence of Mo atoms in their chemical compositions.

The dynamics of variations in the initial
magnetic permeability of alloy MG-3T fabricated from components of
technical purity under the influence of irradiation is somewhat
different (Fig.~1,{\it a}). It is evident that the impurities that enter
into its composition play a substantial role.

In work \cite{4}, we showed that the doping of amorphous alloys
Fe--Si--B with nickel and molybdenum reduces the sensitivity of
their structural factors $i(s)$ to radiation. This fact testifies
that, at least in doped MG-alloys, the influence of radiation on
structure-sensitive magnetic characteristics is not connected with
variations in the short-range integral parameters. In view of the
dependence of radiation-induced changes in the initial magnetic
permeability of MG-alloys on the concentration of doping components,
we may suppose that irradiation brings about the formation of
clusters, which contain, first of all, molybdenum atoms and atoms of
the most mobile alloy components, B and/or Si. The appearance of
such nonmagnetic inclusions can lead to a decrease in the mobility
of domain walls \cite{8}. A possibility of their formation was
confirmed in \cite{9,10}. The regions enriched with molybdenum are
formed in the near-surface layer of amorphous metal alloys (AMAs).
The authors of works \cite{11,12,13,14,15,16,17} considered that the
surface of boron-containing AMAs is also enriched with boron atoms,
especially the free side of a ribbon, which did not contact with the
disk surface at manufacturing an alloy \cite{11,12}. The enhanced
concentration of atoms of this element near the AMA surface was
observed not only in the initial amorphous ribbons, but also in the
thermally treated ones \cite{13,14,15,16,17}. A thermal treatment
additionally increases the boron concentration in the near-surface
layers (about 50~\AA\ in thickness) of ribbons \cite{13}. Therefore,
the formation of clusters containing molybdenum and boron atoms
under the influence of radiation should be expected to take place in
the near-surface layers of a ribbon. The invariance of the initial
magnetic permeability for Fe--Si--B alloys at irradiation doses
above $2.5\times 10^{18}~\gamma
_{\mathrm{quant}}$/$\mathrm{cm}^{\mathrm{2}}$ can be explained by a
saturation of chemical bonds between Mo atoms and atoms of
nonmetallic elements.\looseness=1

%Fig. 2
\begin{figure*}% figure* for wide figure, [h] [!] to change the placement
\includegraphics[width=8.5cm]{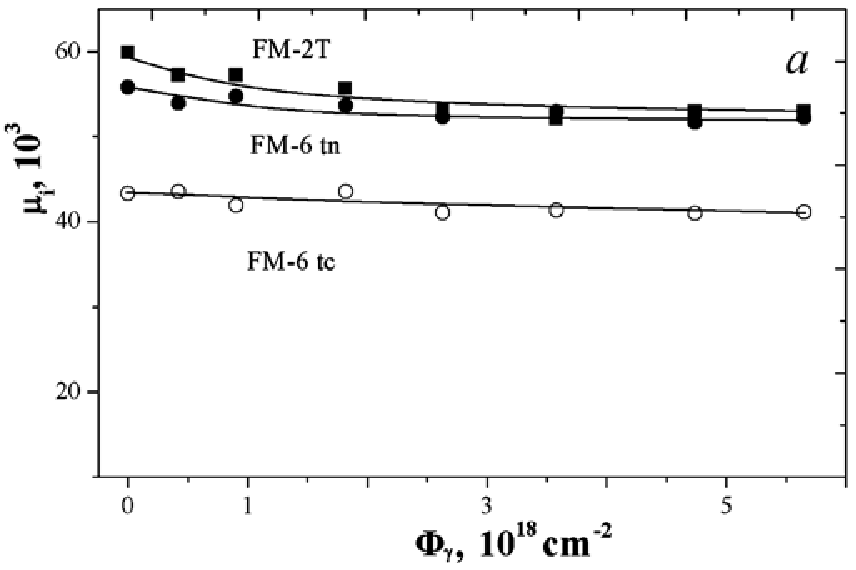}\hspace{0.5cm}\includegraphics[width=8.5cm]{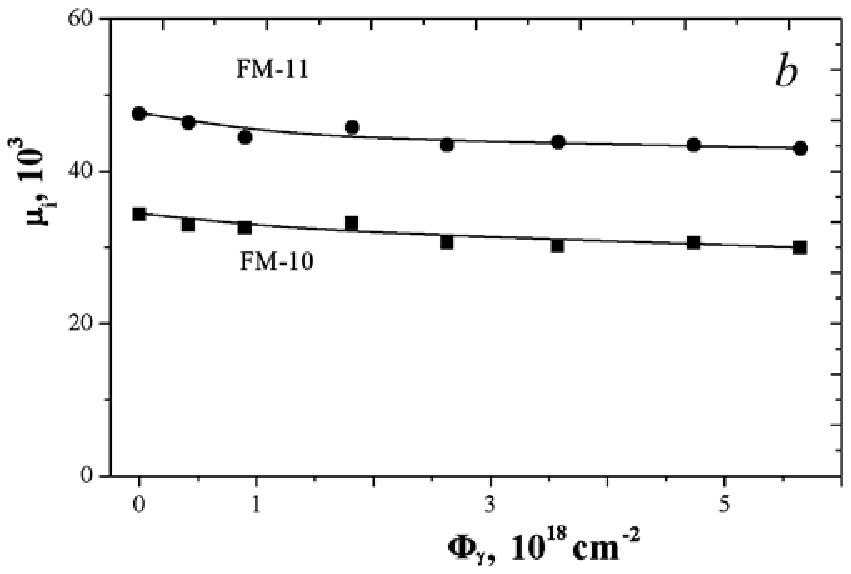}
%\vskip-3mm
\caption{Dose dependences of the initial magnetic permeability
($f=10~\mathrm{kHz}$) of magnetic circuits fabricated from
FM-alloyis  }
\end{figure*}

Additional researches are needed to establish the basic mechanisms of
influence of $\gamma$-radiation on the structure and the magnetic properties of
undoped alloys belonging to the Fe--Si--B system.

In Fig.~2, the dependences of the initial magnetic permeability of
nanocrystalline alloys on the exposure dose of $\gamma $-radiation
are depicted. By comparing the data in Figs.~1 and 2, we arrived at
a conclusion that $\mu _{i}$'s for FM-alloys are less sensitive to
the radiation action, than the magnetic permeabilities of MG-alloys
doped with molybdenum and nickel. The maximum variations of the initial
magnetic permeability stimulated by $\gamma $-irradiation were 12\%
for alloy FM-2T, 8\% for alloy FM-6tn (\textquotedblleft
tn\textquotedblright\ means a thinner ribbon of 24~$\mu
\mathrm{m}$), and 12\% for alloy FM-6tc (\textquotedblleft
tc\textquotedblright\ means a thicker ribbon of 38~$\mu \mathrm{m}$)
(see Fig.~2,\textit{a}). For the magnetic circuits fabricated from FM-10
and FM-11 alloys containing Co, a reduction of $\mu _{i}$ did not
exceed 5 and 7\%, respectively (Fig.~2,\textit{b}).

Nanocrystalline alloys of the Finemet type are biphase systems, in
which both the crystalline and amorphous phases are ferromagnetic at
room temperature. The volume fraction of the crystalline phase in
FM-alloys amounts to about 70\%, and the composition of nanograins
is estimated as Fe--Si (18--23 at.\%) \cite{18,19,20}. Other
elements that enter into FM-alloys are practically insoluble in
$\alpha $-Fe(Si), being located therefore in the amorphous matrix.
In general, variations of the initial magnetic permeability of
FM-alloys stimulated by $\gamma$-radiation can be associated with
atomic reconstructions in the amorphous matrix of alloys, as well as
in the crystal lattice and at the crystal boundaries. However, the
notable variations of the initial magnetic permeability of
crystalline materials can be revealed at much higher doses of
$\gamma $-radiation. This fact is confirmed by our researches
dealing with the influence of the given kind of radiation on
magnetic characteristics of the transformer steel. Since the most
mobile element in the alloy, boron, is contained in the amorphous
matrix, the radiation-induced variations of the initial magnetic
permeability can be associated with its diffusion to crystal
boundaries or with its participation in the formation of
non-magnetic clusters in the amorphous matrix.

In addition, a reduction of the initial magnetic permeability under irradiation
can occur owing to the growth of crystallite dimensions. From
literature data \cite{18,21}, it is known that the increase of a crystal
size from 10 to 40\textrm{~nm} in alloys of the Finemet type worsens the
magnetic properties of the latter. A substantial, by several orders of
magnitudes, reduction of $\mu _{i}$ and an increase of $H_{c}$ take place.
However, with regard for the exposure dose and the fact that
FM-alloys are rather stable quasiequilibrium systems (under their thermal
treatment, the crystalline phase is segregated within a few minutes, i.e.
the fraction of crystalline phase is not changed during almost the whole
annealing process), this mechanism is hardly probable.

As is seen from Figs.~1 and 2, the substantial changes in $\mu _{i}$
and the saturation of the dependences $\mu _{i}(\Phi _{\gamma })$
for MG-alloys doped with nickel and molybdenum, as well as for
FM-alloys, occur within the same interval of $\gamma $-radiation
doses. This fact may mean that the variations of the initial
magnetic permeability stimulated by the radiation treatment
connected with structural modifications in the amorphous matrix of
FM-alloys. Moreover, a similarity of the dose dependences for
initial magnetic permeabilities can also mean that the mechanisms of
radiation influence on a structure of the amorphous matrix in
FM-alloys and on a structure of MG-alloys are qualitatively
identical. The role probably played by molybdenum in MG-alloys can
be characteristic of niobium atoms in nanocrystalline ones. The
formation of non-magnetic inclusions in FM-alloys under the action
of ionizing radiation should be expected to take place at
crystallite boundaries and/or in the near-surface layers of the
ribbon.

\section{Conclusions}

In the present work,

\noindent1)~the doping of amorphous alloys Fe--Si--B with nickel and
molybdenum is found to enhance the sensitivity of their initial magnetic
permeabilities to radiation action; the sensitivity of $\mu_{i}$\ grows with
the Mo concentration and depends a little on the Ni
concentration;

\noindent 2)~the initial magnetic permeabilities of nanocrystalline alloys
on the basis of Fe--Si--B are found to be less sensitive to the action of
$\gamma $-radiation than those for doped amorphous alloys;

\noindent 3)~a hypothesis is advanced that the influence of radiation on
the initial magnetic permeability is associated with the formation of
non-magnetic inclusions in the structure of amorphous alloys and in the
amorphous matrix of nanocrystalline ones.

\rezume{%
ВПЛИВ $\gamma $-ОПРОМІНЕННЯ НА ПОЧАТКОВУ МАГНІТНУ\\ ПРОНИКНІСТЬ
АМОРФНИХ І НАНОКРИСТАЛІЧНИХ\\ СПЛАВІВ НА ОСНОВІ СИСТЕМИ
Fe--Si--B}{В.Ю. Поварчук, В.К. Носенко, А.М. Крайчинський,\\ В.Б.
Неймаш, М.М. Красько, \fbox{В.В. Маслов}} {Методом визначення
фактора індуктивності досліджено залежності початкової магнітної
проникності $\mu _{{i}}$ аморфних і нанокристалічних сплавів на
основі системи Fe--Si--B від дози $\gamma $-опромінення. Виявлено,
що легування аморфних сплавів системи Fe--Si--B молібденом і нікелем
приводить до збільшення радіаційної чутливості їхніх $\mu _{{i}}$.
Початкова магнітна проникність нанокристалічних сплавів на основі
Fe--Si--B менш чутлива до дії $\gamma $-опромінення, ніж легованих
аморфних. Висловлено припущення, що вплив радіації на початкову
магнітну проникність зумовлений утворенням немагнітних включень у
структурі аморфних і в аморфній матриці нанокристалічних сплавів.}

\end{document}